\documentclass[11pt,twoside,a4paper]{article}
\usepackage{latexsym}
\usepackage{graphicx}
\def\be{\begin{equation}}
\def\ee{\end{equation}}
\def\ba{\begin{eqnarray}}
\def\ea{\end{eqnarray}}
\begin{document} 

\title
{\Large {\bf Slow-roll Inflation for Generalized Two-Field Lagrangians}}
\author{Fabrizio Di Marco $^1$\footnote{e-mail: dimarco@bo.infn.it} \, and 
Fabio Finelli 
$^2$\footnote{e-mail: finelli@bo.iasf.cnr.it}
\\
$^1$ Dipartimento di Fisica \\ Universit\`a degli Studi di Bologna
and INFN \\ via Irnerio, 46 -- 40126 Bologna -- Italy \\
$^2$ IASF-BO/INAF \\
Istituto di Astrofisica Spaziale e Fisica Cosmica \\ Sezione di Bologna 
\\ Istituto Nazionale di Astrofisica \\
Via Gobetti 101 -- 
40129 Bologna -- Italy}
\date{\today}
\maketitle
\begin{abstract}
We study the slow-roll regime of two field inflation, in which  
the two fields are also coupled through their kinetic terms. 
Such Lagrangians are motivated by particle physics and by 
scalar-tensor theories studied in the Einstein frame. 
We compute the power spectra of adiabatic and 
isocurvature perturbations on large scales to first order in the 
slow-roll parameters. We discuss the relevance of the extra coupling terms 
for the amplitude and indexes of the power spectra. 
Beyond the consistency condition which involves the amplitude of 
gravitational waves, additional relations may be found in 
particular models based on such Lagrangians:
as an example, we find an additional general consistency 
condition in implicit form for Brans-Dicke theory in the Einstein frame.
%We provide the consistency conditions for scalar-tensor theories 
%for a massless scalaron in the Einstein frame.
%, and we provide the consistency relations for the spectral 
%indeces and amplitudes of the spectra. 
\end{abstract}

\section{Introduction}

A period of inflation in the early Universe explains 
the origin of the large-scale structure by the evolution of initial, 
quantum vacuum 
fluctuations of matter (see \cite{Liddle-Lyth} for a textbook review). 
In the simplest inflationary model 
the dynamics is driven by a {\em single} scalar field, whose quantum fluctations 
produce a primordial, scale invariant spectrum for curvature perturbations. 
The amplitude of curvature perturbation remains constant on super-horizon scales until the 
time when the perturbation re-enters into the Hubble scale, 
when the universe is dominated by radiation or matter.

As soon as more than one matter field is considered during inflation, isocurvature 
perturbations may arise among different component and may also affect the curvature perturbations on 
large scales. Isocurvature fluctuations arise naturally when two or more scalar fields 
slow-roll during inflation \cite{Sjetp,Pol,MS}.

A useful formalism which splits the original two-field dynamics 
in a tangential and orthogonal (to the 
trajectory in phase space) basis has been developed \cite{GWBM}: 
this method results in a straightforward 
identification of curvature and isocurvature fluctuations at first order in 
perturbation theory, whose evolution equations are  
regular during inflation. This formalism has been further developed for general 
theories with two scalar fields \cite{GNVT,NR,DFB,Kosh}. 

In this paper we study the two-field slow-roll regime for theories 
described by the following action
\be
S = \int d^4 x \sqrt{-g} \left[ \frac{R M_{\rm pl}^{2}}{2} - 
\frac{g^{\mu \nu}}{2}
\partial_\mu \varphi \partial_\nu \varphi - 
\frac{e^{2 b (\varphi)}}{2} g^{\mu \nu}
\partial_\mu \chi \partial_\nu \chi - V (\varphi, \chi) \right]
\label{action}
\ee
where $M_{\rm pl}=1/\sqrt{8 \pi G}$ is the reduced Planck mass. 
The non-standard kinetic term for $\chi$ appears in $\sigma$-model theories or 
in scalar-theories for gravity after a transformation to the conformal 
Einstein frame \cite{dicke,shinji}. In a previous paper \cite{DFB} the 
splitting of adiabatic 
and isocurvature perturbations for the action (\ref{action}) was studied.
It was found that the extra term generated by the coupling in kinetic term for $\chi$, couples adiabatic and 
isocurvature modes for scales larger than Hubble radius also for scaling 
solutions \cite{DFB}. A stronger correlation of adiabatic and isocurvature 
modes is therefore expected for $b_\varphi \ne 0$. 
%The main consequence is the presence of correlations between adiabatic and isocurvature 
%perturbations also for the case when the ratio of the kinetic terms of the two fields is 
%costant, and imply relevant modification in all the observational quantities to respect 
%to the case of two field ordinary inflation.

It is interesting to investigate if there are generic predictions for inflationary models 
in which the dynamics is not driven by a single field \cite{BMR}. For 
$b_\varphi=0$ \footnote{The case with $b \ne 0$ and 
$b_\varphi = 0$ is trivial since this constant can be included in a redefinition of $\chi$.} 
in Eq. (\ref{action}), the only model-independent prediction is a consistency 
relation among the tensor to scalar ratio, the gravity waves spectral index and the cross-correlation 
between curvature and isocurvature perturbations \cite{WBMR}, which modifies the single field consistency relation.
The present paper is devoted to the prediction of inflationary models 
based on the action (\ref{action}) for $b_\varphi \ne 0$.

The outline of the paper is as follows. 
In section II we review the formalism and the main equations obtained in 
\cite{DFB} 
for the theory in Eq. (\ref{action}). In section III we study the slow-roll 
approximation. In section IV we study the dynamics of curvature and isocurvature 
perturbations during inflation and in section V we give the final power spectra. 
%In section VI we calculate explicity the evolution of the 
%perturbations during the slow-roll regime, and we discuss the relevance of the extra-term. 
%In section V we write the power spectra at the beginning of the post-inflationary era, 
%using the formalism of the transfer function, and in section VI 
%we derive the general spectral indexes and the consistency relations. 
In section VI we focus on some model dependent relations and in section VII we apply 
our results to scalar-tensor theories studied in the Einstein frame.
We conclude in section VIII.    
%derive the spectral indices formulae and the consistency relation. 
%These sections generalize the work for minimally coupled scalar fields 
%with standard kinetic terms of Gordon et al. \cite{Bartolo}......

\section{Basic Equations}
In this first section we shall review the equations of motion deriving from 
(\ref{action}). Such equations can be also found in  \cite{DFB}, 
but we feel to rewrite them here in order to make our paper self-contained. 
The equations of motion for the two homogeneous field are:
\be
\ddot \varphi + 3 H \dot \varphi + V_\varphi = b_\varphi
e^{2 b} \dot \chi^2 \,,
\label{backphi}
\ee
\be
\ddot \chi + (3 H + 2 b_\varphi \dot \varphi) \dot \chi 
+ e^{- 2 b} \, V_\chi = 0 \,,
\label{backchi}
\ee
and the Einstein equations are: 
\be
%H^2 = \frac{1}{3 M_{\rm pl}^2} \left[ \frac{1}{2} \dot \varphi^2 + \frac{e^{2 b}}{2} \dot \chi^2 + V \right] \,,
H^2 = \frac{8 \pi G}{3} \left[ \frac{1}{2} \dot \varphi^2 + \frac{e^{2 b}}{2} \dot \chi^2 + V \right] \,,
\ee
\be \label{dot_H}
%\dot H = - \frac{1}{2 M_{\rm pl}^2} \left[ \dot \varphi^2 +e^{2 b} \dot \chi^2\right] \,. 
\dot H = - 4 \pi G \left[ \dot \varphi^2 +e^{2 b} \dot \chi^2\right] \equiv - 4 \pi G \dot\sigma^2 \,,
\ee
where the last equation is not independent from the others. 
The average and orthogonal fields are \cite{DFB}:
\be
d \sigma = \cos \theta \, d \varphi + \sin \theta
\, e^b \, d \chi \, ,
\label{adiabaticfield}
\ee
\be
ds = %\frac{e^b}{\dot \sigma} [ \, \dot \varphi \, d \chi - \dot \chi \, d \varphi \, ] = 
e^b \cos \theta \, d \chi 
- \sin \theta \, d \varphi  \, ,
\ee
with:
\begin{eqnarray}
\cos \theta &=& \frac{\dot \varphi}{\sqrt{\dot \varphi^2 +
e^{2 b} \dot \chi^2}}\, , \nonumber \\
\sin \theta &=& \frac{e^b \dot \chi}{\sqrt{\dot \varphi^2 +
e^{2 b} \dot \chi^2}}\, .
\end{eqnarray}
The average field $\sigma$ and the angle $\theta$ satisfy, respectively:
\be
\ddot \sigma + 3 H \dot \sigma + V_{\sigma} = 0 \,,
\ee
\be
\dot \theta = - \frac{V_s}{\dot \sigma} - b_\varphi \dot \sigma \sin \theta \, , 
\ee
where:
\be \label{Vsig}  V_{\sigma} =  V_\varphi \,\cos\theta + e^{- b}\, V_\chi \, \sin \theta\, , \ee
%and we define also:
\be \label{Vs}  V_s = - V_\varphi\sin \theta + e^{- b} \,  V_\chi\cos \theta\, \,. \ee
We then pass to equations for the fluctuations \cite{DFB}. By using the longitudinal gauge for the metric fluctuations:
\be ds^2 = -(1 + 2 \Phi)dt^2 + a^2(1 - 2 \Phi)d{\bf x}^2, \ee
the equation for the average Mukhanov variable $Q_{\sigma} = \delta\sigma + \frac{\dot\sigma}{H}\Phi $ is:
\begin{eqnarray} 
 &\ddot Q_\sigma& +\; 3H \dot Q_\sigma  + \left[ \frac{k^2}{a^2} 
+ V_{\sigma \sigma} + \dot \theta \frac{V_s}{\dot \sigma} 
- \frac{1}{M_{\rm pl}^2 a^3} \left(\frac{a^3  \dot\sigma^2}{H}\right)^{.}
- b_\varphi \dot \varphi 
\frac{V_\chi e^{-b}}{\dot 
\sigma} \sin \theta
\right] Q_\sigma \nonumber 
\\ &=& - 2 (\frac{V_s}{\dot \sigma}\delta s)^. + 2 \left( 
\frac{V_\sigma}{\dot 
\sigma} + \frac{\dot H}{H} \right) \frac{V_s}{\dot \sigma} \delta s\, , 
\, \label{adiab}
\end{eqnarray}
and for $\delta s$ we have:
\begin{eqnarray}
\ddot{\delta s} &+& 3 H \dot{\delta s} + \left[ \frac{k^2}{a^2} 
+ V_{ss} + 3 \dot \theta^2 - b_{\varphi \varphi} \dot \sigma^2 +
b_\varphi^2 g (t) + b_\varphi f (t) \right]
\delta s \nonumber \\
&=& - \frac{k^2}{a^2} \frac{\Phi}{2 \pi G} \frac{V_s}{\dot \sigma^2} 
\, ,\label{entropy}
\end{eqnarray}
where:
\begin{eqnarray}
g (t) &=& - \dot \sigma^2 (1 + 3 \sin^2\theta)\, , \nonumber \\ 
f (t) &=& V_\varphi (1 + \sin^2 \theta) - 4 V_s \sin \theta \, .
\end{eqnarray}

We note again that Eq. (\ref{adiab}) has the correct single field limit 
and all the equations of this paragraph 
agree with those in \cite{GWBM} when $b_\varphi=0$.

\section{Slow-Roll Expansion}
%In this section we generalized the formalism of slow-roll approximation for the inflation to the system drived from action (\ref{action}). The equation of motion for the fields $\varphi$ and $\chi$ which we obtain from the variation of this action are:
Under the assumption of the slow-roll for both fields $\varphi$ and $\chi$ the equations of motions at first-order are:
\be
\dot \varphi =\dot\sigma\cos\theta \simeq -\frac{ V_\varphi}{3 H}\, , \hspace{1.3cm}
\dot \chi =\dot\sigma \sin\theta e^{-b} \simeq -\frac{V_\chi}{3 H} e^{-2b}\, ,\label{condizslowroll}
\ee
\be
H^2(\varphi,\chi) \simeq \frac{8 \pi G}{3} V(\varphi,\chi)\, .
%\quad \quad \epsilon \equiv - \frac{\dot H}{H^2}
\label{H^2p}
\ee
We note that we do not keep the viscous term $2 b_\varphi \dot \varphi 
\dot \chi$ in Eq. (\ref{backchi}) 
to lowest order in a slow-roll expansion. 
%By defining the global slow-roll parameter $\epsilon \equiv - \frac{\dot H}{H^2}$, the other slow-roll 
By defining the slow-roll parameters as:
\be
\epsilon \equiv - \frac{\dot H}{H^2} \simeq \frac{M_{\rm pl}^2}{2}\left(\frac{V_\sigma}{V}\right)^2\, ,
\ee
\be \label{epsilon}
\epsilon_{\varphi} =\frac{M_{\rm pl}^2}{2}\left(\frac{V_\varphi}{V}\right)^2 \simeq \epsilon \cos^2 \theta\, ,
\;\;\;\; \epsilon_{\chi} =\frac{M_{\rm pl}^2}{2}\left(\frac{ V_\chi}{V}\right)^2 e^{-2b} \simeq \epsilon \sin^2 
\theta \, ,
\ee
%\epsilon_{\varphi} =\frac{1}{16 \pi G}\left(\frac{ V_\varphi}{V}\right)^2\;\;\;\;\;\epsilon_{\chi} =\frac{1}{16 \pi G}\left(\frac{ V_\chi}{V}\right)^2 e^{-2b}\ee
\be \eta_{\varphi \varphi} = M_{\rm pl}^2 \frac{V_{\varphi \varphi}}{V}\, ,\;\;\;\;\;\eta_{\varphi \chi} = M_{\rm pl}^2  \frac{V_{\varphi \chi}}{V}e^{-b}\, ,\;\;\;\;\;\eta_{\chi \chi} = M_{\rm pl}^2 \frac{V_{\chi \chi}}{V} e^{-2b}\, ,
%\be \eta_{\varphi \varphi} = \frac{1}{8 \pi G} \frac{V_{\varphi \varphi}}{V}\;\;\;\;\;\eta_{\varphi \chi} = \frac{1}{8 \pi G}  \frac{V_{\varphi \chi}}{V}e^{-b}\
%\;\;\;\;\eta_{\chi \chi} = \frac{1}{8\pi G} \frac{V_{\chi \chi}}{V} e^{-2b}
\label{parametri1}
\ee
$$\;\;$$
\be \epsilon_b = 8 M_{\rm pl}^2 b^2_\varphi \, , %\eta_b = 8M_{\rm pl}^2 \,
\ee
%\be \label{parametri2}\epsilon_b = \frac{b^2_\varphi }{\pi G}\hspace{1.5cm} \eta_b =\frac{b_{\varphi \varphi}}{\pi G} \label{parametri2} \ee
%where we have introduced the reduced Planck mass $M_{\rm pl}^2 = (8 \pi G)^{-1}$. 
%We note that $b_{\varphi \varphi}$ appears only to second order in slow-roll parameters and therefore we neglect it.
neglecting the other terms in the equations (\ref{backphi}) 
and (\ref{backchi}):
\begin{eqnarray} \left \{
\left|\ddot \varphi\right|,\left|b_{\varphi}e^{2b}\dot\chi^2\right|\right\}\; &\ll& \;\left\{\left|V_{\varphi}\right|,3\left|H\dot\varphi\right|\right\}\, , \\
\left\{\left|\ddot \chi\right|e^{2b},\left|b_{\varphi}\dot\varphi\dot\chi\right|e^{2b}\right\}\; &\ll&\;\left\{\left| V_{\chi}\right|, 3\left|H\dot\varphi\right|\right\}, \end{eqnarray}
we have the conditions for 
the slow-roll: $ \epsilon_{i} \ll 1,\;\;|\eta_{ij}| \ll 1\;\;\forall\;i,j = {\varphi, \chi}$ and $\epsilon_{b} \ll 1 $.
These conditions for the parameters (\ref{epsilon}) and (\ref{parametri1}) are the generalization of 
the slow-roll conditions, and the one for $\epsilon_b$ arises directly 
by requiring that  $\varphi$ and $\chi$ slow-roll. %, and the condition for $\eta_b$ by requiring that: $ \{\dot\epsilon_{i}, \;\;|\dot\eta_{ij}| \}=  O(\epsilon^2_{i},\eta_{ij}^2)$. 
Now we can to extend the formalism of average and entropy field to the slow-roll parameters and we define: 
\be \eta_{\sigma \sigma} \equiv \eta_{\varphi \varphi} \cos^2 \theta + \eta_{\varphi\chi} \sin 2 \theta + 
\eta_{\chi\chi} \sin^2 \theta = \frac{V_{\sigma\sigma}}{3 H^2} \label{etasigma} \, ,\ee
\be \eta_{\sigma s}\equiv (\eta_{\chi\chi} - \eta_{\varphi \varphi}) \sin\theta\cos\theta + \eta_{\varphi\chi} 
(\cos^2\theta-\sin^2\theta) = \frac{V_{\sigma s}}{3 H^2}\, , \label{etasigmas}\ee
\be  \eta_{ss}\equiv  \eta_{\varphi \varphi} \sin^2\theta - \eta_{\varphi\chi} \sin 2 \theta + \eta_{\chi\chi} 
\cos^2\theta = \frac{V_{ss}}{3 H^2}\,  . \label{etas} \ee
With these slow-roll conditions:
$$ \epsilon_{i} \ll 1\,\;\;|\eta_{ij}| \ll 1\;\;\;\;\forall\;i,j = {\sigma, s}, \quad \quad \epsilon_b \ll 1\, ,$$
the background slow-roll solution is:
\be
\dot \sigma^2 \simeq \frac{2}{3} \epsilon V \, , \quad \quad 
\frac{\dot \theta}{H} \simeq - \eta_{\sigma s} + \frac{1}{2} {\rm 
sign}(b_{\varphi}){\rm sign} {\left(\frac{V_{\chi}}{V}\right)}\sqrt{\epsilon_b \epsilon_\chi} \cos^2 \theta \, ,
\ee
\be
\frac{\ddot \sigma}{H \dot \sigma} \simeq \epsilon - 
\eta_{\sigma \sigma} + \frac{1}{2}
{\rm sign}(b_{\varphi}){\rm sign}
\left(\frac{V_{\chi}}{V}\right)\sqrt{\epsilon_b\epsilon_\chi} \sin 
\theta \cos \theta \, ,
\ee
and the equations of motion for $Q_\sigma$ and $\delta s$ on large scales 
%the adiabatic and isocurvature perturbations (\ref{adiab}) and (\ref{entropy}) 
become:
\be \dot Q_\sigma = A H Q_\sigma + B H\delta s \, , \label{SRQ2}\ee 
\be \dot\delta s =  H S \delta s \, , \label{SRs2}\ee
where:
\begin{eqnarray} 
A (\epsilon_i,\eta_{ij}) &=& - \eta_{\sigma\sigma} + 2\epsilon + \frac{1}{2}{\rm sign}(b_{\varphi}){\rm 
sign}{\left(\frac{V_{\chi}}{V}\right)}
\sqrt{\epsilon_b\epsilon_\chi}\sin\theta\cos\theta\, , \label{A} \\
B (\epsilon_i,\eta_{ij}) &=& - 2 \eta_{\sigma s} - 
{\rm sign}(b_{\varphi}){\rm sign}
{\left(\frac{V_{\chi}}{V}\right)}\sqrt{\epsilon_b 
\epsilon_\chi} \sin^2 \theta \nonumber \\
&=& 2\;\frac{\dot\theta}{H} -
{\rm sign}(b_{\varphi}){\rm sign}
{\left(\frac{V_{\chi}}{V}\right)}\sqrt{\epsilon_b
\epsilon_\chi}\, ,
\label{B} \\
S(\epsilon_i,\eta_{ij}) &=& - \eta_{ss} -\frac{1}{2} {\rm sign}(b_{\varphi}){\rm sign}
{\left(\frac{V_{\varphi}}{V}\right)}
\sqrt{\epsilon_b\epsilon_\varphi} (1 + \sin^2 \theta ).
\label{Z} 
\end{eqnarray} 
%and $\pm$ discriminate the sign of the ratio $V_\varphi / V $, $V_\chi / V$ or of the funcion $b(\varphi)$.
On large scales the entropy field perturbations evolve
independently of the adiabatic field,
but in contrast to the $b_\varphi=0$ case isocurvature perturbations do
affect curvature perturbations
also when $\eta_{\sigma s} = 0$ (or $\dot \theta=0$).
From Eqs. (\ref{Z}) it is clear that we need to know two more
parameters, $\theta$
and $\epsilon_b$, with respect to the four needed in the case with
$b_\varphi=0$: this means that the original asymmetry
between $\varphi$ and $\chi$ cannot be completely hidden by
the diagonalization in $\sigma$ and $s$.

\section{Evolution of Fluctuations During Inflation}
 
Following \cite{GWBM} we write at Hubble crossing the amplitude of 
perturbations as: 
\be Q_{\sigma}|_{(k=a_* H_*)} = \frac{H_*}{\sqrt{2k^3}}e_{\sigma}(k)\, ,
\hspace{1.2cm}
\delta s |_{(k=a_* H_*)}=\frac{H_*}{\sqrt{2k^3}}e_{s}(k)\, , \label{Qzero}\ee
where $H_*$ is the Hubble parameter evaluated at horizon crossing, and the random variables 
$e_{\sigma}(k)$ and $e_{s}(k)$ satisfy:
\be \langle e_I(k)\rangle=0\; \hspace{0.5cm} 
\hbox{and}\hspace{0.5cm}\;\langle e_I(k)\; \bar e_J(k') 
\rangle = \delta_{IJ}\delta(k-k')
\hspace{0.6cm}\{I,J\}=\{\sigma,s\} \label{random}.
\ee
It is important to stress that the above 
Eqs. (\ref{Qzero},\ref{random}) imply that adiabatic and isocurvature 
fluctuations have same spectrum and amplitude and vanishing 
correlation {\em at horizon crossing}. At 
the end of this section we shall elaborate more on this assumption. 

Integrating the (\ref{SRs2}) and supposing that the change of 
$S$ is negligible during inflation, 
after the substituing the second formula of (\ref{Qzero}) 
we find that the isocurvature perturbations evolve as: 
\be \delta s(t) = \frac{H_*}{2k^3} e^{S(N_* - N(t))} e_s(k) \label{evols}\ee
where $N_* = \int^{t_F}_{t_*} H dt\,  $ corresponds to the number of the e-folds between the horizon crossing and the end of inflation.The slow-roll solution for $Q_{\sigma}$ is:
\be  Q_{\sigma} (t) = \frac{H_*}{2k^3}e^{A(N_* - N(t))}
e_{\sigma}(k) + \frac{H_*}{2k^3} e^{S(N_* - N(t))}e_s(k)\, . \ee
These formulae allow to calculate the power spectra:
\be \langle  Q_{\sigma}(k)\; \bar Q_{\sigma}(k')  \rangle = \frac{2\pi^2}{k^3}{\cal P}_{Q_{\sigma}}\delta(k-k')\, ,\hspace{0.6cm}\langle \delta s(k)\; {\bar\delta s}(k')\rangle = \frac{2\pi^2}{k^3}{\cal P}_s\delta(k-k')\, , \ee
and the equation:
\be \langle  Q_{\sigma}\;\bar{\delta s} \rangle \equiv \frac{2\pi^2}{k^3}{\cal C}_{Q_{\sigma} s}\delta(k-k') \ee
 defines the {\it correlation} between the variables $Q_{\sigma}$ and $\delta s$.
%The calculate gives:
%\be {\cal P}_{Q_{\sigma}} = \frac{H_*^2}{(2\pi)^2}e^ {2 A (N_*-N(t))}\left[(1 + \left(\frac{B}{\gamma}\right)^2 
%(1- e^{-\gamma (N_*-N(t))})^2 \right]\ee 
%mentre quello delle perturbazioni di isocurvatura vale:
%\be
%{\cal P}_s = \frac{H_*^2}{(2\pi)^2}e^ {2 S (N_*-N(t))}
% \ee
%\be {\cal C}_{Q_{\sigma} s} = \frac{H_*^2}{(2\pi)^2}\frac{B}{\gamma}e^{2 S(N_*-N(t))}\left(e^{\gamma 
%(N_*-N(t))}-1\right) \label{corrfunctiont}\ee
%where $\gamma = A - S$.

Now we would like to express the results in terms of curvature and isocurvature fluctuations, defined as:
\begin{eqnarray}
\zeta = H \frac{Q_{\sigma}}{\dot\sigma}\, , \quad \quad {\cal S} = H 
\frac{\delta s}{\dot\sigma} \,,
\end{eqnarray}
which are related by \cite{DFB}:
\begin{eqnarray}
\dot\zeta &=& \frac{H}{\dot H}\frac{k^2}{a^2} \Phi + \frac{2H}{\dot\sigma} 
\dot\theta \delta s + 2 b_{\varphi} H \sin\theta \delta s \nonumber \\
& = & \frac{H}{\dot H}\frac{k^2}{a^2} \Phi - 2 \frac{V_s}{\dot\sigma} 
{\cal S}\, ,
\end{eqnarray}
and whose power spectra at Hubble crossing are given by:
\be
{\cal P}_{\zeta}|_* \; \simeq \;{\cal P}_ {\cal S}|_* \;\simeq \;\frac{1}{(2 \pi)^2}\frac{H_*^4}{\dot\sigma_*^2} \, . \label{Spectrain}\ee
In terms of these quantities the relevant power spectra at the end of 
inflation are:
\begin{eqnarray} {\cal P}_{\zeta} &=&  \frac{H^2_*}{(2\pi)^2} \frac{1}{2 
M_{\rm pl}^2 \epsilon_*}
\left[1 + \left(\frac{B}{\gamma}\right)^2(1- e^{-\gamma N_*})^2 \right]\, , 
\label{PR}\\
\label{PS} {\cal P}_ {\cal S} &=&  \frac{H^2_*}{(2\pi)^2} 
\frac{1}{2 M_{\rm pl}^2 \epsilon_*}e^{-2 \gamma N_*} \, ,\\
%\end{eqnarray}
 \label{CORR} {\cal P}_ {\cal C} &=& {\cal C}_{\zeta {\cal S}} =  
\frac{H^2_*}{(2\pi)^2}\frac{1}{2 M_{\rm pl}^2 \epsilon_*}\frac{B}{\gamma}
e^{-2 \gamma N_*}(e^{\gamma N_*} - 1 )\, \end{eqnarray}
where:
\be
\gamma = A - S \,. 
\label{gamma}
\ee
We note 
that despite appearance the limit for $\gamma \rightarrow 0$ (in which 
isocurvature perturbations are not damped) is well defined.  
The power spectrum of gravitational waves is:
\be {\cal P}_T =
{\cal P}_T|_*
= \frac{8}{M_{\rm pl}^2} \frac{H^2_*}{(2\pi)^2}
\label{PT}
\ee
and remains unchanged after horizon crossing and through the radiation 
era. The spectral indexes are defined as: 
\be 
n_m -1 \equiv  \frac{d \ln {\cal P}_m}{d \ln k} \quad \quad
m=\zeta,{\cal S},{\cal C} \,,
\quad n_T \equiv \frac{d \ln {\cal P}_T}{d \ln 
k} \,.
\label{index_def}
\ee
Once evaluated at horizon crossing these spectral indexes are:
\begin{eqnarray}
n_\zeta - 1 |_* &=& n_{\cal S} - 1 |_*  
\nonumber \\
& = & - 6 \epsilon_* + 2 \eta_{\sigma \sigma *} - 
{\rm sign}(b_{\varphi}){\rm sign}
\left(\frac{V_{\chi}}{V}\right)\sqrt{\epsilon_{b \, *} \epsilon_{\chi \, 
*}} \sin \theta_* \cos \theta_* \, ,
\nonumber \\
n_T |_* &=& -2 \epsilon_* \, ,
\end{eqnarray}
and $P_{\cal C} |_* = 0$, i. e. adiabatic and isocurvature fluctuations 
are considered uncorrelated at Hubble crossing as from Eqs. 
(\ref{Qzero},\ref{random}).

Eqs. (\ref{Qzero},\ref{random}) mean that the variables to quantize (or 
randomize in terms of classical numbers) are $Q_\sigma, \delta s$ and 
their spectra and amplitude are the same at horizon crossing, although 
their evolution equations are different and coupled, as is clear from 
Eqs. (\ref{adiab},\ref{entropy}). 
It is conceivable that this assumption may be violated in certain models 
\cite{shinji2} for $b_\varphi=0$. 
We therefore expect that a non-vanishing correlation at horizon 
crossing may be present also for $b_\varphi \ne 0$.
It is also conceivable to quantize the Mukhanov variables 
$Q_\varphi$, $Q_\chi$ (associated to $\varphi$, $\chi$, respectively) 
instead of $Q_\sigma, \delta s$. Since
\be
Q_\sigma = \cos \theta \, Q_\varphi + \sin \theta \, e^b \, Q_\chi\, ,  \quad 
\quad 
\delta s = - \sin \theta \, Q_\varphi + \cos \theta \, e^b \, Q_\chi \, ,
\ee 
by imposing Eq. (\ref{random}) for $I,J=Q_\varphi \,, Q_\sigma$ one has 
\be
P_{\cal C} |_* = \frac{H^2_*}{\dot \sigma^2_*} \sin \theta_* \cos 
\theta_* \left( e^{2 b_*} \langle Q_\chi^2 
\rangle - \langle Q_\varphi^2 \rangle \right) \, .
%\nonumber \\
%&=& \frac{H^3_*}{4 \pi^2 \dot \sigma^2_* k^3} \sin \theta_* \cos
%\theta_* \left( e^{-2 b_*} - 1\right) \,,
\ee
It is then clear
that adiabatic and isocurvature modes are really uncorrelated at
horizon crossing when one of the two field dominates also for 
$b_\varphi \ne 0$. If Eq. (\ref{Qzero}) is used for $Q_\varphi$ and  
$\langle Q_{\chi \, k}^2 \rangle = H^2_* e^{-2 b_*}/(2 
k^3)$ \cite{Garcia-Bellido:1995fz} in order to take into account the 
extra damping term in the equation of motion for $\chi$, the correlation 
is again zero also for the $b_\varphi \ne 0$ case. 
When both fields are important at horizon crossing, a 
correlation may be present if the two Mukhanov variables do not have the 
same amplitude at horizon crossing - up to rescaling.

\section{After Inflation}

The end of inflation depends strongly on the form of the potential $V$. 
As an example, let us focus on the case $V(\varphi, \chi) = 
e^{-\beta \varphi/M_{\rm pl}} {\tilde V} (\chi)$, where $\beta > 0$. 
%(such potential is generic after conformal transformations). 

The exact scaling solution found in \cite{FB2001} 
(and discussed in more detail in \cite{DFB}) for ${\tilde V}$ independent on $\chi$ is a 
threshold regime between the domination of $\varphi$ and $\chi$. 
The field $\chi$ may end up not oscillating even when ${\tilde V} (\chi)$ 
is convex depending on $b(\varphi)$. In such a case inflation may be ended 
by instant preheating \cite{instant} or by symmetry breaking in a third 
field, like in hybrid models (but in this last case our formalism 
is not sufficient).

In the case in which inflation ends by $\chi$ oscillations 
($\varphi$ may oscillate as well) one should study if parametric 
amplification of scalar perturbations occurs during the preheating 
phase \cite{preh,FB}. 
Indeed, the fact that both $\varphi$ and $\chi$ 
slow-roll during inflation garantees that 
a mixture of adiabatic and isocurvature perturbations with similar 
infrared spectra is present at the end of inflation: this is one of the 
necessary conditions for having an amplification of curvature 
perturbations during preheating \cite{FB}. In the worst case 
(as for a quartic potential 
potential for the inflaton coupled to a second field by a 
dimensionless parameter $g^2$ \cite{preh,FB}) fluctuations may grow until 
the non-linear stage even on large scales \cite{nonlinear} and 
the scenario would not be compatible with our universe. This fine tuning 
for the inflaton coupling parameters leads to the 
conclusion that a simple quartic potential 
is under theoretical, and not only observational, pressure 
\cite{quarticwmap}.

In a scenario compatible with observations fluctuations should remain small, 
although they may change drastically after inflation is ended. 
On large scales, adiabatic and 
isocurvature fluctuations evolve according to:
\be \dot \zeta =  \alpha(t) H(t) {\cal S},\label{eqzeta}\ee
\be \dot {\cal S} = \delta(t) H(t) {\cal S} \,.
\label{eqS} 
\ee
%where $\alpha$ and $\delta$ are time-dependent functions.  
By integrating over time we can apply the formalism of the 
transfer matrix \cite{WBMR, AGWS} in order to study how the correlation 
between adiabatic and isocurvature perturbations builds up:
\be
{\zeta(t)\choose{\cal S}(t)}\quad = \left(\matrix{ 1 & T_{{\zeta}{\cal 
S}}\cr 0 & T_{{\cal S}{\cal S}}\cr}\right) 
{\zeta(t_*)\choose{\cal S}(t_*)}\quad
\ee
where:
\begin{eqnarray}
T_{{\cal S}{\cal S}} (t_*, t) &=& \exp\left( \int_{t_*}^t \delta (t') 
H(t') d t' \right)\, , \nonumber \\  
T_{{\zeta}{\cal S}} (t_*, t) &=& \int_{t_*}^t \alpha(t') H (t') T_{{\cal 
S}{\cal S}} 
(t_*, t') d t' \,.
\end{eqnarray}
We note that we are implicitly assuming that 
$T_{{\zeta}{\cal S}}  \,, T_{{\cal S}{\cal S}}$ depend on $k$ only through $t_*$, 
the instant at which fluctuations leave the Hubble radius. Such assumption 
is useful also for models compatible with observations, in which fluctuations are amplified 
during (p)reheating, but in a $k$-independent way in the region $k \sim 0$ (where the fluctuations 
relevant for observations are located during (p)reheating).
The power spectra are therefore:
\begin{eqnarray}
{\cal P}_{\zeta} &=& (1 + T_{\zeta{\cal S}}^2) 
\, {\cal P}_ \zeta|_* = {\cal P}_ \zeta|_* ( 1 + \cot^2 \Delta )\,,
\label{gen_spectra_R}\\\;
\nonumber \\
{\cal P}_{\cal S} &=& T_{{\cal S}{\cal S}}^2 \, {\cal P}_ \zeta|_* \, ,
\label{gen_spectra_S} \\\;\nonumber\\
{\cal C}_{\zeta{\cal S}} &=& T_{\zeta{\cal S}}T_{{\cal S}{\cal S}}
{\cal P}_ \zeta|_* \, ,  \label{gen_spectra_C}
\end {eqnarray}
where the measure of the correlation is introduced as 
the {\it cross-correlation angle} $\Delta$: 
\be
\cos \Delta = \frac{P_{\cal C}}{\sqrt{{\cal P}_ {\zeta} {\cal P}_ {\cal
S}}} = \frac{T_{\zeta{\cal S}}}{\sqrt{{\cal P}_ {\zeta} {\cal P}_ {\cal
S}}}\, ; \label{Delta}
\ee
$\Delta$ allows to reconstruct curvature perturbation spectrum 
at horizon crossing: 
%if it is known at the end of inflation:
\be{\cal P}_ {\zeta}|_* \simeq {\cal P}_ {\zeta} \sin^2\Delta . 
\label{Pold}\ee

We note that with the relation $\tan \Delta = \gamma e^{\gamma 
N_*}/(B(e^{\gamma 
N_*}-1))$ Eqs. 
(\ref{gen_spectra_R},\ref{gen_spectra_S},\ref{gen_spectra_C})
agree with Eqs. (\ref{PR},\ref{PS},\ref{CORR}), the formalism of transfer 
functions may be also used during inflation leading to the correct 
results.

%Formulae (\ref{PR}), (\ref{PS}) and (\ref{CORR}) generalize the results 
%find in \cite{BMR} to the case in which 
%$\chi$ have a noncanonical kinetic term, and reduct to (46)-(48) of 
%\cite{Bartolo} for $\epsilon_b$ = 0.

The spectral indexes defined as in Eqs. (\ref{index_def}) are: 
\begin{eqnarray}
n_{\zeta} - 1 &=& - 6\epsilon + 4 \epsilon (\cos\Delta)^2 + 
2 \eta_{\sigma \sigma} (\sin\Delta)^2 +
4  \eta_{\sigma s}\sin\Delta \cos\Delta \nonumber\\
&& + 2 \eta_{ss} (\cos\Delta)^2 
+ 2\;{\rm sign}(b_{\varphi}){\rm sign}\left(\frac{V_{\chi}}{V}\right)
\sqrt{\epsilon_b\epsilon_\chi} (\sin\theta)^2 \sin\Delta \cos\Delta \nonumber\\
&& +{\rm sign}(b_{\varphi}){\rm sign}
\left(\frac{V_{\varphi}}{V}\right)\sqrt{\epsilon_b\epsilon_\varphi}
(1 + \sin^2 \theta) \cos^2 \Delta \nonumber\\
& & - {\rm sign}(b_{\varphi}){\rm sign}
\left(\frac{V_{\chi}}{V}\right) 
\sqrt{\epsilon_b\epsilon_\chi} \sin\theta \cos\theta 
\sin^2 \Delta \, , \nonumber \label{nr} \\
%\;&& \nonumber\\
\label{ns} n_{\cal S} -1 &=& -2\epsilon -2 S 
= - 2\epsilon + 2 \eta_{ss} +{\rm sign}(b_{\varphi}){\rm sign}
\left({\frac{V_{\varphi}}{V}}\right) 
\sqrt{\epsilon_b \epsilon_\varphi}(1+(\sin\theta)^2)\, , \nonumber \\
\;&& \nonumber\\
n_{\cal C} - 1 &=&  - 2\epsilon + 2 \eta_{ss} + 2 \eta_{\sigma s}\tan\Delta +{\rm sign}(b_{\varphi}){\rm sign}
\left({\frac{V_{\varphi}}{V}}\right)\sqrt{\epsilon_b \epsilon_\varphi}[1 +(\sin\theta)^2 ]
\nonumber\\
&& + {\rm sign}(b_{\varphi}){\rm sign}\left({\frac{V_{\chi}}{V}}\right)\sqrt{\epsilon_b \epsilon_\chi}
(\sin\theta)^2 \tan\Delta \nonumber\\
 &=& n_{\cal S} -1 + \left(2 \eta_{\sigma s} +{\rm sign}(b_{\varphi}){\rm sign}\left({\frac{V_{\chi}}{V}}\right)
\sqrt{\epsilon_b \epsilon_\chi}(\sin\theta)^2 \right)\tan\Delta  \, ,
\label{nc} \nonumber \\
n_T &=& - 2 \epsilon \, ,
\label{nT}
\end{eqnarray}
where we have omitted to indicate that the slow-roll parameters are evaluated at 
Hubble crossing (as thereafter in the paper). The crucial assumption of explicit 
$k$-independence of the transfer functions has allowed to derive the final spectral 
indexes just in terms of the slow-roll parameters at Hubble crossing and the 
correlation angle $\Delta$, since $\alpha_*=B$ and $\delta_*=-\gamma$.

As already said, the power spectrum of gravitational waves 
remains unchanged after horizon crossing and we find the consistency 
condition:
\be 
\frac{{\cal P}_T}{{\cal P}_{\zeta}} =
- 8 n_T \left(1 - \frac{{\cal C}_{ {\zeta}{\cal S}}^2}
{ {\cal P}_{\zeta} {\cal P}_{\cal S}}\right) \label{relcon}
\ee
as for the case of double inflation with $b_\varphi=0$ \cite{BMR,WBMR}. 
As expected this relation does not change for $b_\varphi \ne 0$, but it 
becomes an upper bound in presence of additional fields \cite{WBMR}.

\section{Model-Dependent Relations}

The class of inflationary models studied here 
contains two more parameters than usual double inflation with 
$b_\varphi=0$ \cite{BMR, WBMR}. The amount of parameters is therefore $9$: 
6 inflationary parameters plus the Hubble scale during inflation plus two 
transfer functions. The number of input parameters for observations always 
remain $8$ to first order in 
slow-roll expansion: $4$ spectra plus $4$ spectral indexes. 
In a situation where the parameters are more than the "observables", the 
relation for gravitational waves which persists with respect to the 
$b_\varphi=0$ case - although expected - is a benefit. Therefore we need to look for 
model dependent relations.

We have already noticed in the introduction that isocurvature and 
adiabatic perturbations are coupled also for scaling solution, 
i. e. when $\dot \theta =0$ (since $V_s$ is not simply proportional to 
$\dot \theta$ \cite{DFB}). 
This coupling is also evident in the definition of the $B$ 
parameter in Eq. (\ref{B}), which is not vanishing also for 
$\eta_{\sigma s} \sim 0$. Therefore, perturbations can be effectively 
decoupled when 
$\varphi$ dominates, but never when $\chi$ dominates. 
As a consequence, $n_{\cal C} \ne n_{\cal S}$ despite $\eta_{\sigma s} 
\sim 0$.

We conclude this section observing that, for the curvaton case \cite{LW}, where
$$ \sin\Delta \sim 0 $$ 
and the adiabatic perturbation at horizon crossing 
is negligible, we find that tensor perturbations 
are negligible (${\cal P}_T \simeq 0$) and a relation among the scalar indices:
\begin{eqnarray}  
n_{\cal \zeta} \simeq  n_{\cal C} \simeq n_{\cal S} 
&=& 1 - 2\epsilon + 2 \eta_{ss} + {\rm sign}(b_{\varphi}){\rm sign}
\left(\frac{V_{\varphi}}{V}\right) \sqrt{\epsilon_b\epsilon_\varphi}
[1+(\sin\theta)^2] \nonumber \\
&=& 1 - 2\epsilon + 2 \eta_{ss} + 2 b_{\varphi} M_{\rm pl} \sqrt{2 \epsilon} \cos \theta
[1+(\sin\theta)^2] 
\label{same} 
\end{eqnarray}  
which is qualitatively similar to the case with $\epsilon_b = 0$ - the 
three spectral indexes are the same - but quantitatively different. All 
the curvaton phenomenology is therefore changed if $\epsilon_b \ne 0$ and $\varphi$ was not 
negligible during inflation.

\section{Application to Scalar-Tensor Theories in the Einstein 
Frame}

We now apply our study to the particular case 
of scalar-tensor theories studied in the Einstein frame. 
The analysis of the consistency conditions in the Jordan frame is in progress.
For the case of a 
massless dilaton ($\varphi$) the potential in the action (\ref{action}) 
is:
\be V(\varphi,\chi) = e^{4 b(\varphi)} U (\chi) \,.
\label{Brans-Dicke}
\ee
The case of Brans-Dicke cosmology is 
obtained for
$b (\varphi) \propto \varphi$ \cite{BD} and will be discussed in Sec.
$(7.3)$. Starting from a scalar-tensor theory in the Jordan frame
\be
S = \int d^4 x \sqrt{-\tilde g} \left[ \frac{\tilde R}{16 \pi} F(\phi) -
G(\phi) \frac{\tilde g^{\mu \nu}}{16 \pi}
\partial_\mu \phi \partial_\nu \phi -
\frac{\tilde g^{\mu \nu}}{2} 
\partial_\mu \chi \partial_\nu \chi - U (\chi) \right] \,,
\label{action_st}
\ee
the action (\ref{action}) with the potential (\ref{Brans-Dicke}) is 
recovered by rewriting (\ref{action_st}) in the conformal frame with metric $g_{\mu \nu} 
= \tilde g_{\mu \nu} (G F(\phi))$ (the function 
$b(\varphi)$ which parametrizes the relation between the original dilaton $\phi$ and 
its conformal one $\varphi$). 

With respect to other multi-field 
inflationary theories, scalar-tensor cosmologies may follow a simple 
evolution: the inflaton $\chi$ decays in matter fields after inflation and 
$\varphi$ evolves coupled to the matter trace until the present 
time, determining the coupling constants. For the background 
Eqs. (\ref{backphi},\ref{backchi}) are simply rewritten 
as:
\begin{eqnarray}
\ddot \varphi + 3 H \dot \varphi &=& b_\varphi T_\chi\, , \nonumber \\
\dot \rho_\chi + 3 H (\rho_\chi + p_\chi ) &=&  - b_\varphi \dot \varphi 
T_\chi \, , 
\end{eqnarray}
where the $\chi$ energy density is $\rho_\chi = e^{2 \, b} \dot \chi^2/2 + 
e^{4 b} U(\chi)$ and $T_\chi = -\rho_\chi + 3 p_\chi$ is the 
$\chi$ 
trace. Adiabatic perturbations were studied for the case in 
Eq. (\ref{Brans-Dicke}) neglecting the correlation between 
adiabatic and isocurvature modes \cite{Garcia-Bellido:1995fz}.

%This application is motivated by the aim of investigate the evolution of perturbations for the extendend 
%chaotich inflation when both the Brans-Dicke scalar field and the inflaton slow-roll.\par
The first consequence of the choice (\ref{Brans-Dicke}) is the reduction from six 
to {\em four} independent slow-roll parameters:
\be \epsilon_{\varphi} = 8 M^2_{\rm Pl} b_{\varphi}^2 \label{par_1}\, ,\ee
\be \epsilon_{\chi} = \frac{ M^2_{\rm Pl}}{2}\left(\frac{U_{\chi}}{U}\right)^2 e^{-2b} \label{par_2} \, ,
\ee 
\be \eta_{\varphi \varphi} = 4 M^2_{\rm Pl} ( b_{\varphi\varphi} + 4  b_{\varphi}^2) = 4 M^2_{\rm Pl}   b_{\varphi\varphi} + 2 \epsilon_{\varphi} \label{par_3} \, ,\ee
\be \eta_{\chi \chi} = M^2_{\rm Pl}\left( \frac{U_{\chi\chi}}{U}\right) 
e^{-2b} \, , \label{par_4} \ee
and two which are not independent: 
\be \epsilon_b=\epsilon_\varphi\, , \quad \quad 
\eta_{\varphi \chi} = 2  {\rm sign}(b_{\varphi}){\rm sign} 
\left(\frac{U_{\chi}}{U}\right) \sqrt{\epsilon_{\varphi}\epsilon_{\chi}} 
\,.
\ee

\subsection{Extended inflation: $\sin \theta \sim 0$}

If we suppose that the evolution of the massless dilaton dominates, 
$\sin\theta << \cos\theta \,$ ($\epsilon_\chi << \epsilon_\varphi$), we 
have:
\be A= - 4  M^2_{\rm Pl} b_{\varphi \varphi} + 2 \epsilon_{\chi}\, , \hspace{0.5cm} B = -4{\rm sign}(b_{\varphi}){\rm sign} \left(\frac{U_{\chi}}{U}\right) \sqrt{\epsilon_{\varphi}\epsilon_{\chi}}\, , \hspace{0.5cm}\ee
\be  S= -\eta_{\chi\chi} - \frac{1}{2} \epsilon_{\varphi}. \ee
The isocurvature spectral index is:
\be n_{\cal S} -1 = - \epsilon_{\varphi} + 2 \eta_{\chi \chi}\, . \ee
When $\cos\Delta = 0$ the adiabatic spectrum coincides with Eq. (4.18) of \cite{Garcia-Bellido:1995fz}, 
and the spectral index of the 
adiabatic perturbation 
\be n_{\zeta} - 1 = - 2 \epsilon_{\varphi} + 8   M^2_{\rm Pl} b_{\varphi \varphi} \ee
agrees with \cite{Garcia-Bellido:1995fz}. 

\subsection{Chaotic inflation: $\cos \theta \sim 0$ }

If the inflaton $\chi$ dominates, $\cos \theta << \sin \theta$ 
($\epsilon_\varphi << \epsilon_\chi$) and we have:
\be A= -\eta_{\chi \chi} + 2 (\epsilon_{\chi} + \epsilon_{\varphi})\, , \hspace{0.4cm} B =  3{\rm sign}(b_{\varphi}){\rm sign} \left(\frac{U_{\chi}}{U}\right) \sqrt{\epsilon_{\varphi}\epsilon_{\chi}}\, , \hspace{0.4cm}\ee
\be  S= -3 \epsilon_{\varphi} - 4 M^2_{\rm Pl} b_{\varphi \varphi} \, , \ee
and the isocurvature index is:
\be   n_{\cal S} -1 = - 2 \epsilon_\chi + 2 \eta_{\varphi \varphi} = 
- 2 \epsilon_{\chi} + 4 \epsilon_{\varphi} + 8  M^2_{\rm Pl} b_{\varphi \varphi}. \ee
In the absence of correlation  ($B = 0$ and $\cos\Delta = 0$) the spectral index of adiabatic perturbation is:
\be n_{\zeta} - 1 = - 6\epsilon_{\chi} + 2 \eta_{\chi \chi} \ee
which coincides with the Eq. (5.7) of  \cite{Garcia-Bellido:1995fz} if one poses $\alpha_* = 0$ and corrects the factor $e^{2a_*} $. This implies that the growth of the factor $b$ reduces the variation of the tilt.\par

\subsection{Brans-Dicke Theory}

The Brans-Dicke theory \cite{BD} was throughly investigated in the 
past \cite{SY,Garcia-Bellido:1995fz}: it can be obtained by $F(\phi) = 
\phi$ and $G (\phi) = \omega/\phi$, with $\omega$ as a $\phi$ 
independent parameter. In such 
way one obtains $2 b (\varphi) = - \varphi/(\sqrt{\omega+3/2} {M_{\rm 
pl}})$. For $b (\varphi) \propto \varphi$ we have {\em three} 
slow-roll parameters 
since $\eta_{\varphi \varphi} = 2 \epsilon_\varphi$.
Therefore it is possible to make one prediction in addition to Eq. 
(\ref{relcon}). We have checked that in the 
simplest case of a quadratic and quartic potential for $\chi$ there is no
amplification during preheating \footnote{During preheating the
splitting used here in $Q_\sigma$ and $\delta s$ is not so useful because
the evolution equations become singular. It is more useful to use the
Mukhanov variables associated to $\delta \varphi$ and $\delta \chi$ as
in \cite{FB}.}. 
It is interesting to 
note that for a quartic potential, $\varphi$ displays 
oscillations to leading order, too. Indeed, the $\chi$ background time 
average energy density redshift like radiation, but the oscillations in the 
$\chi$ trace drive $\varphi$. Any field coupled to the dilaton may then be 
excited by parametric resonance. However, we take $\Delta$ in order to 
take our results completely general. 

\begin{figure}
%\vspace{3cm}
\begin{tabular}{cc}
 \includegraphics[scale=0.58]{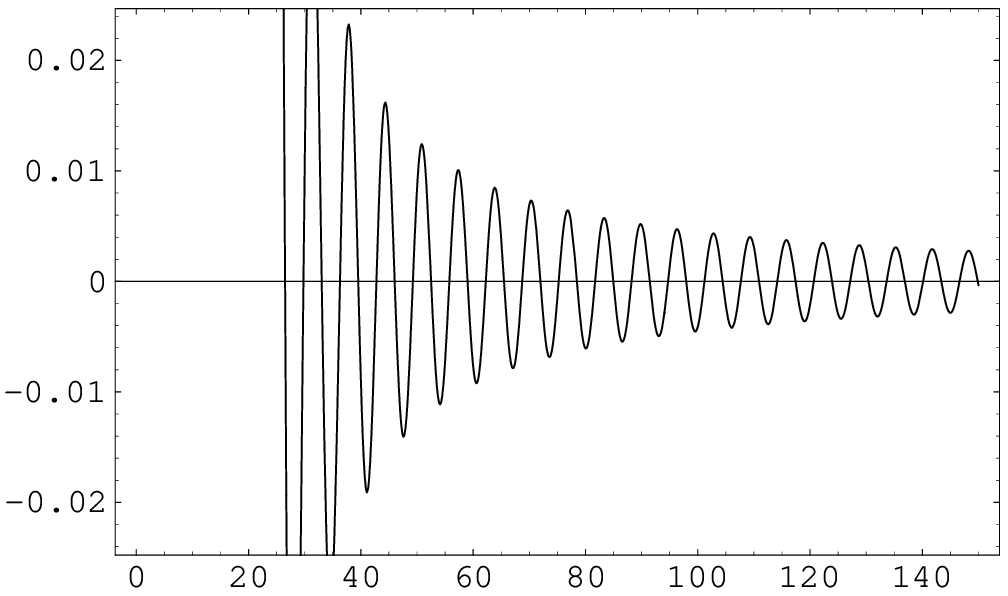}&
 \includegraphics[scale=0.58]{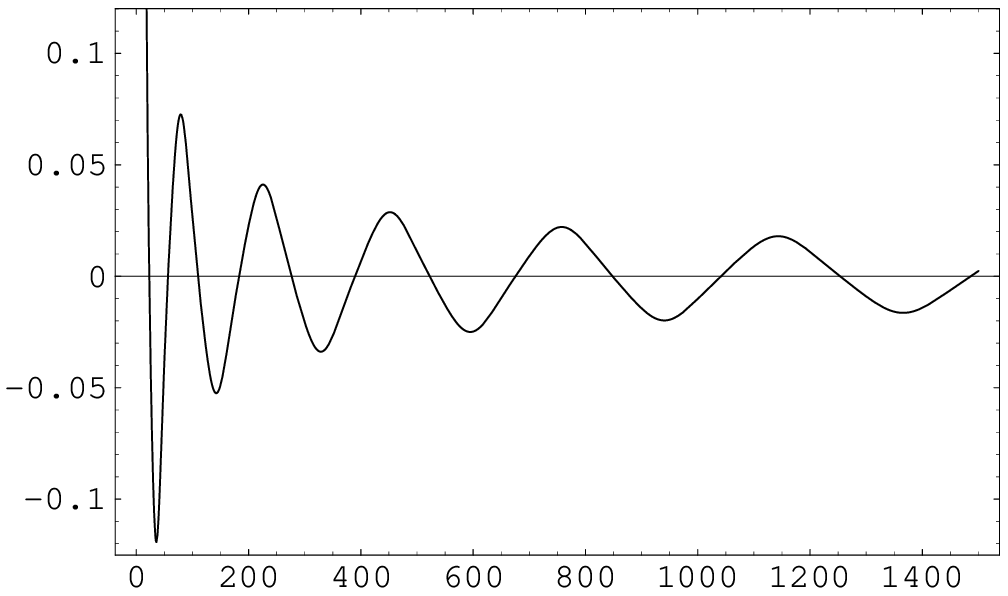}\\
 \includegraphics[scale=0.58]{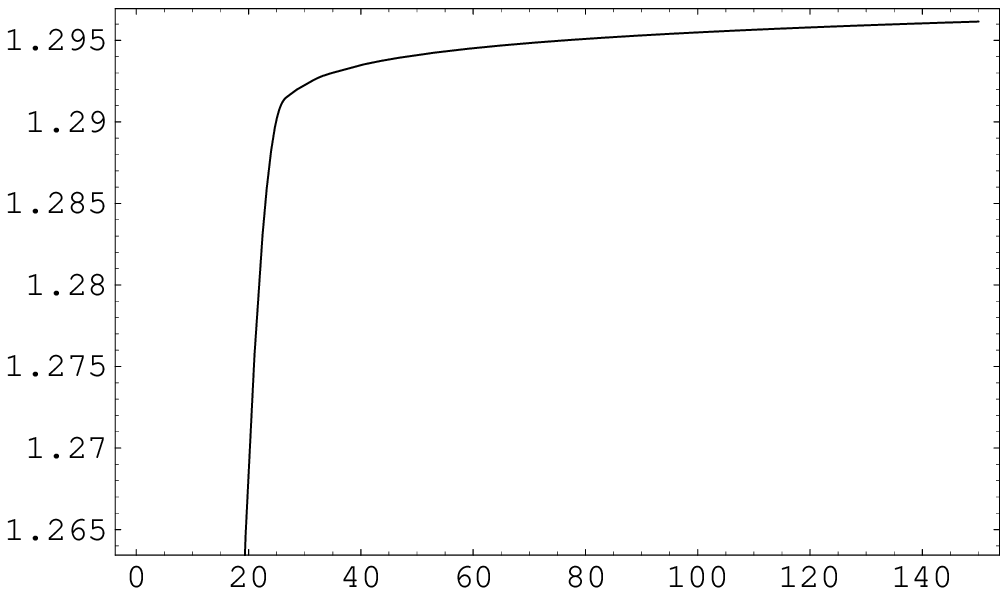}&
 \includegraphics[scale=0.58]{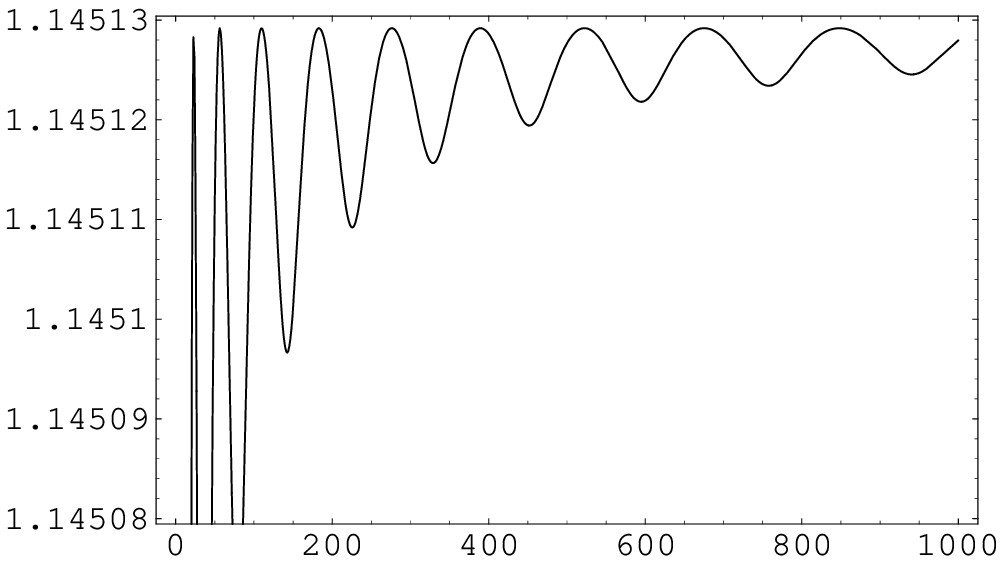}
\end{tabular}

\caption{Evolution of $\chi (t)$ (top) and $\varphi (t)$ (bottom) for 
\protect{}$U(\chi) = m^2 \chi^2/2$ (left) and \protect{}$U(\chi) = \lambda \chi^4/4$ 
(right). The fields are in \protect{}$1/\sqrt{G}$ units and the 
$x$-axis is $m t$ for the massive case and 
\protect{}$\sqrt{\lambda/G} t$ for the self-interacting case. After a slow-roll regime, 
$\varphi$ oscillates as well in the case of a self-interacting potential for 
$\chi$ bottom panel, to the right).}

\end{figure}

We therefore obtain in implicit form the following consistency condition 
among the four spectral indexes and the correlation angle $\Delta$. An 
intermediate step is to give the relation for $\eta_{\chi \chi}$:
\be
2 \eta_{\chi \chi} = \frac{n_\zeta - 1}{\sin^2 \Delta} + n_{\cal S} - 1
- \cot^2 \Delta ( 2 n_{\cal C} - n_{\cal S} - 1 )
%\nonumber \\ & & 
- \frac{n_T}{2} ( 8 - 5 \cos^2\theta )\,
,\ee
where we have used $n_T = - 2 \epsilon$. Plugging this relation in the 
relation for the spectral indexes we get:

\be
\left \{ \begin{array} {ll}

(n_s -1)(\sin\Delta)^2(\sin\theta)^2 &= n_T(\sin\Delta)^2(1 - \cos^2\theta\sin^2\theta -2\cos^2\theta) \\
&+ ({n_{\zeta} -1})\cos^2\theta  - (\cos{\Delta})^2 (2 n_c - n_s -1)\cos^2\theta
 \\
\;\;\\
(n_c - n_s)\sin\Delta\cos\Delta
&= [(n_{\zeta} + n_s -1)
- (\cos{\Delta})^2 (2 n_c -1)]\cos\theta\sin\theta 
\\
 & - (\sin\Delta)^2\cos\theta\sin\theta[1+ 
 \frac{n_T}{2}(3 + 2\sin^2\theta)] \, .\label{relcon2}
\end{array}\right.
\;\;
\ee
As expected on the number of parameters, the latter system of equations in terms of $\theta$ 
is the additional consistency condition in implicit form. We note that for 
$\sin \Delta \sim 0$ the above relation is a subcase of Eq. (\ref{same}).

We stress that Brans-Dicke theory - studied in the Einstein frame - is just an example of 
the possibility to have a consistency condition in addition to Eq. (\ref{same}). Other 
physical models based on Einstein gravity may display the same interesting feature.

\section{Conclusions}

We have studied double field inflationary models based on the action 
(\ref{action}). As already pointed out previously \cite{DFB} the 
correlation of isocurvature and curvature perturbations is strengthened by the 
non-standard kinetic term for $\chi$.

We have computed the power spectra for adiabatic and isocurvature perturbations 
to first order in the slow-roll parameters, 
taking into account their correlation which builds up after fluctuations 
leave the Hubble radius. Our approach was limited by three assumptions.
First, we consider slow-roll parameters to lowest order, although 
there are several efforts to go beyond this approximation \cite{WKB}: on the 
other hand, this approximation allows us to take into account the 
correlation between adiabatic and isocurvature perturbations to lowest 
order. Second, we consider adiabatic and isocurvature fluctuations 
uncorrelated at horizon crossing. 
Third, we have considered the transfer functions 
independent on $k$, apart from the dependence on the instant in which 
fluctuations leave the Hubble radius. 
Within these usual assumptions, our results are completely general.

The search for inflationary consistency conditions to first order in 
slow-roll parameters for theories based on Eq. (\ref{action}) 
seems desperate. Indeed, the parameters are in general more than the 
"observables", and the consistency condition in Eq. (\ref{relcon}) remains 
only because tensor and scalar modes are decoupled. However, 
for some physical models in which there are just three independent slow-roll 
parameters an additional prediction should be present: 
Brans-Dicke theories - studied in the Einstein frame - is just an 
example in which the consistency condition in Eq. (\ref{relcon2}) among 
the four spectral indexes and the 
correlation angle holds. On the basis of numbers of parameters, a  
consistency condition in addition to Eq. (\ref{relcon2}) is also expected 
in other class of scalar-tensor theories and may 
become an important theoretical predictions. Therefore, in some inflationary 
models based on string theory, spatial 
variation of the coupling constants (given by fluctuations in the dilaton) are 
correlated with density fluctuations in a way which is predictable.

It is important to note that the observational relevance of isocurvature 
perturbations of massless moduli is strictly connected to scalar-tensor 
theories to first order in perturbation theory. Indeed, for $b_\varphi=0$ 
the background density of an uncoupled 
massless scalar is completely washed out after inflation: because of the 
$\dot \varphi$ fast redshift dependence ($\propto a^{-3}$), the  
observational relevance of such perturbation is washed out in the later 
evolution. On the opposite, 
with $b_\varphi \ne 0$, $\varphi$ is coupled both to the 
inflaton in the early universe and to non-relativistic matter, 
becoming a non-negligible ingredient of the primordial soup at late times.

We conclude by adding that dilaton isocurvature 
perturbations may be relevant for CMB and LSS observations, both in 
the massless or effectively massive case. 
Comparing to other nearly massless mode, such as  
radiation-quintessence \cite{AF,isode}, dilaton isocurvature perturbations 
may be more relevant for observations because of the dilaton coupling to non-relativistic 
matter.

\section{Acknowledgements}

We thank Robert Brandenberger for suggestions and comments on the manuscript.

%\vspace{1cm}

%{\bf References}

\end{document}